\documentclass{article}
\usepackage{graphicx} 
\usepackage[export]{adjustbox}
\usepackage{natbib}
\usepackage{amsmath}
\usepackage{authblk}

\newcommand\mancha{\textsc{Mancha3D-2F~}}

\title{Mixing, heating and ion-neutral decoupling induced by Rayleigh-Taylor instability in prominence-corona transition regions.}

\author[1]{V. S. Lukin}
\author[2,3]{E. Khomenko}
\author[4]{B. Popescu Braileanu}

\affil[1]{National Science Foundation, Alexandria, VA, USA}
\affil[2]{Instituto de Astrof\'{\i}sica de Canarias, La Laguna, Tenerife, Spain}
\affil[3]{Departamento de Astrof\'{\i}sica, Universidad de La Laguna, La Laguna, Tenerife, Spain}
\affil[4]{Centre for mathematical Plasma Astrophysics, KU Leuven, Leuven, Belgium}

\date{January 2024}

\begin{document}

\maketitle

\begin{abstract}

This study explores non-linear development of the magnetized Rayleigh-Taylor instability (RTI) in a prominence-corona transition region.  Using a two-fluid model of a partially ionized plasma, we compare RTI simulations for several different magnetic field configurations.  We follow prior descriptions of the numerical prominence model \citep{rti1,rti2,rti3} and explore the charged-neutral fluid coupling and plasma heating in a two-dimensional mixing layer for different magnetic field configurations.  We also investigate how the shear in magnetic field surrounding a prominence may impact the release of the gravitational potential energy of the prominence material.  We show that the flow decoupling is strongest in the plane normal to the direction of the magnetic field, where neutral pressure gradients drive ion-neutral drifts and frictional heating is balanced by adiabatic cooling of the expanding prominence material.  We also show that magnetic field within the mixing plane can lead to faster non-linear release of the gravitational energy driving the RTI, while more efficiently heating the plasma via viscous dissipation of associated plasma flows.  We relate the computational results to potential observables by highlighting how integrating over under-resolved two-fluid sub-structure may lead to misinterpretation of observational data. 

\end{abstract}

\section{Introduction}
\label{sec:intro}

Cool plasma condensations in the solar corona, such as solar prominences or coronal rain, are formed from partially ionized plasma. These structures are typically thermally insulated from the corona by the magnetic field, as highly anisotropic heat conduction in coronal conditions acts along the field lines. Between the cool and dense condensation and the hot and rarefied corona lies a narrow transition layer often termed as the prominence-corona transition region (PCTR). Within this layer, the parameters transition from strongly coupled, weakly ionized plasma at the core of the structures to fully ionized but weakly collisionally coupled plasma in the coronal part. The thermal and dynamic structure of the PCTR is believed to be determined by several processes, such as turbulent mixing, radiative cooling, and heat conduction.

In observations, the prominence-corona transition region is detected from space using EUV HeII 304 \AA\ resonance line or transition-region spectral lines, which are emission lines from heavier elements but in low ionization states compared to the corona \citep{Parenti2014}. The width of spectral lines observed in the PCTR exhibits signatures of non-thermal motions \citep{Stellmacher+etal2003, Parenti+Vial2007}, which are frequently interpreted as signatures of waves, microturbulence, or unresolved fine structures.

Cooling, linked to mixing during the nonlinear phase of the Kelvin-Helmholtz instability (KHI), was proposed by \cite{Hillier+Arregui2019} as a mechanism that establishes the temperature within the PCTR. These researchers concluded that turbulent mixing, rather than heating (specifically turbulent heating in their case), primarily influences the eventual temperature of this layer. This cooler layer, with an intermediate temperature, may provide a conducive environment for increased radiative losses, further cooling, and condensations \citep{Hillier+etal2023}.

One crucial process within PCTR is the interaction between the ionized and neutral components of the plasma, which is the focus of our current study. The structures under discussion in this paper are significantly smaller than typical prominences. Nonetheless, these cool coronal structures are embedded within their own PCTR-like regions. Recently, \cite{MartinezGomez+etal2022} demonstrated the existence of large velocity decoupling, accompanied by frictional heating and temperature enhancement, at the transition region surrounding coronal raindrops. The rapid change in plasma parameters across coronal raindrops' narrow surface layers facilitates the decoupling between charged and neutral plasma components.
 
Similar behavior has been documented in a series of our papers investigating plasma dynamics associated with the Rayleigh-Taylor instability (RTI) in solar prominence threads \citep{rti1,rti2,rti3}. In these studies, we extensively explore how both the linear and nonlinear progression of RTI depends on the prominence mass load, the supporting magnetic field structure, the ion-neutral collision coupling strength, as well as viscosity and ionization-recombination reaction rates. Our findings reveal that due to imbalanced ionization/recombination (considered under coronal approximation, as discussed in  \cite{2012Meier}), secondary structures easily form alongside an over-ionized material layer in regions surrounding falling drops. We identified ion-neutral collisions as the key factor influencing the development or damping of these small-scale structures. In our modeling, we observed charge-neutral flow decoupling and cross-field plasma motions within the PCTR. Furthermore, the presence of field shear occasionally led to the dynamic formation, breakup, and merging of coherent 2.5D magnetic structures and plasmoid sub-structures.
 
A significant disparity exists between simulations of this nature and what current observations, facilitated by our best instrumentation, can detect. In observations, dynamic processes within optically thin plasma are projected onto the plane of the sky. However, the spatial resolution, limited either by instrument capabilities or atmospheric seeing conditions, hinders the detection of the finest structures. Numerous authors have endeavored to identify potential dynamical decoupling between plasma components by measuring sets of ionized and neutral spectral lines targeting optically thin prominence plasma \citep{2016Khomenko, 2017Anan, Wiehr2019, Wiehr2021, Zapior2022, GonzalezManrique2023}. 
Despite inherent uncertainties in these observations, straining the sensitivity limits of the instrumentation used, all studies, except \cite{2017Anan}, have reported a slight excess in the amplitude of ion variations over neutrals. These observations employ stringent selection criteria for optically thin plasma, necessitating the argument that the same plasma volume is sampled by both ionized and neutral species. However, it's noteworthy that the analyzed areas often coincide with the borders of prominence structures. Therefore, it is plausible to speculate that part of these signals is emitted by the prominence-corona transition region, though it is crucial to bear in mind that the spectral lines used in these studies measure plasma at lower temperatures more typical of chromospheric conditions.

In this study, our focus lies in investigating the non-linear mixing occurring during the later phases of the Rayleigh-Taylor instability (RTI) at the border of simulated prominence threads. Building upon the analysis presented in prior works \citep{rti1,rti2,rti3}, we delve deeper into understanding the energy flow and the heating mechanisms within the thread-corona transition layer and the overall volume affected by various mechanisms present in our simulations. These mechanisms include compressional effects, frictional heating resulting from charges-neutral flow decoupling, as well as contributions from viscosity, ionization, and magnetic field dissipation. To achieve this, we utilized simulations with four times higher resolution compared to those in \cite{rti1,rti2}. This increase in resolution proved pivotal in adequately resolving the physical processes, overcoming the intrinsic dissipation limitations of our numerical code \citep{Popescu+etal2018}. Additionally, we endeavored to bridge the gap between observed and simulated plasma properties by deliberately degrading our simulations to a coarser resolution. This allowed us to establish a limit to the observed drifts and enabled a discussion on potential misinterpretation of observational data due to unresolved two-fluid substructures.

\section{Overview of the simulations}
\label{sec:sim_summary}
\subsection{Numerical model}

The simulation results presented below were obtained using the \mancha code.  The \mancha code numerically solves the two-fluid equations described in \cite{rti3} and \cite{Popescu+etal2018}.\footnote{See Section~\ref{sec:heating} below for the numerical values of the collisional transport coefficients used in the simulations.}

In this paper, we compare simulations with three different prominence thread initial condition setups, each of which has been described in our prior publications \citep{rti1,rti2,rti3}.  The simulations are conducted in the $\hat{x}\times\hat{z}$ plane, with vector quantities allowed to have all three spatial dimensions, and the only difference among the three initial conditions is the spatial profile of the shear in the magnetic field supporting the prominence material.

For clarity and completeness, we repeat the description of the initial condition designed as an RTI-unstable initial equilibrium, first formulated in \cite{Leake2014}:
\begin{align} 
\rho_{\rm c0} &= m_H n_{\rm i0} = m_H n_0 \exp{(-z / H_c)}, \label{eqs:rti_setup_ni} \\
\rho_{\rm n0} &= m_H n_{\rm n0} = m_H \left[ \frac{n_{\rm n00}}{\cosh^2{(2 z / L_0 - 1)}} + n_{\rm nb} \right], \label{eqs:rti_setup_nn} \\
T_0 &= T_b \frac{\cosh^2{(z / L_0 - 0.5)}}{(z / L_0 - 0.5)^2 + L_{\rm t}}, \label{eqs:rti_setup_T}  \\ 
p_{\rm n0} &= n_{\rm n0} k_{\rm B} T_0\,, \nonumber  \\
p_{\rm c0} &= 2 n_{\rm i0} k_{\rm B} T_0\,, \\
B_0 &= B_{00}  \Bigg\{1 + \beta_p \bigg[\frac{n_{\rm i0}}{n_0} \Big(1-\frac{T_0}{T_b}\Big) - 0.5 \frac{n_{\rm n0} T_0}{n_0 T_b} \nonumber \\ 
 & -\frac{1}{H_c n_0} \big(0.5 L_0  n_{\rm n00} \tanh{(2z/L_0 - 1}) + n_{\rm nb}  z \big)  \bigg]\Bigg\}^{0.5}, \label{eqs:rti_setup_B1} \\
\mathbf{B}_0 &= B_0 ( \hat{x} \sin{\theta} + \hat{y} \cos{\theta}) \text{, with } \theta(z) \equiv \theta_0\frac{\pi}{180^\circ} \tanh{(z / L_{\rm s})} \,, \label{eqs:rti_setup_B2}
\end{align}
where the ion number density at $z=0$ is set to $n_0$ = $10^{15}$~m$^{-3}$; the peak neutral number density at $z=L_0/2$ is $n_{\rm n00}$ = $10^{16}$~m$^{-3}$; the background temperature of the corona is $T_b$ = 2.02$\times$10$^5$~K; the neutral number density corresponding to the temperature of the corona is $n_{\rm nb}$ = 3.5$\times$10$^9$~m$^{-3}$; the background magnetic field intensity is $B_{00}=10^{-3}$~T; and the charges gravitational 
scale height is $H_{\rm c} = 2 {k_{\rm B} T_b}/{(m_{\rm H} g)}$. The characteristic length scale is set as $L_0=1$~Mm, and the plasma $\beta_p = {2 n_0 k_{\rm B} T_b}/({B_{00}^2}/2\mu_0)$ is calculated at $z=0$ using $B_{00}$ as the value of the magnetic field to give  $\beta_p \approx$ 1.4$\times$10$^{-2}$. The plasma temperature profile given by Eq.~\ref{eqs:rti_setup_T} has the value of $L_{\rm t} = 20$ chosen so that the temperature of the prominence thread is closest to observed values and the ionization fraction 
$\xi_{\rm i} = {\rho_{\rm c}}/{(\rho_{\rm c} + \rho_{\rm n})}$ = 0.091  remains small \citep[see][]{Leake2014}.

The initial magnetic field configuration is defined by Eqs.~\ref{eqs:rti_setup_B1}--\ref{eqs:rti_setup_B2} and in all three cases is initially contained in the $\hat{x}\times\hat{y}$ plane, with the variation only in the $\hat{z}$-direction.  The three initial magnetic field configurations considered is this paper are 
\begin{description}
    \item P:   $\theta_0 = 0^\circ$, with the magnetic field purely in $\hat{y}$-direction normal to the plane of the simulation \\
    \item L1:   $\theta_0 = 1^\circ$ and $L_s = L_0 = 1$~Mm, with the magnetic field sheared on a scale longer than the prominence density gradient scale \\
    \item L2:   $\theta_0 = 1^\circ$ and $L_s = L_0/2 = 0.5$~Mm, with the magnetic field sheared on the same scale as the prominence density gradient scale
\end{description}
\noindent
The P and L2 configurations have been previously considered in \cite{rti1,rti2}, while simulations initialized in the L1 configuration have been discussed in \cite{rti1,rti3}.  

In addition to the initial equilibrium, for all simulations presented in this paper a small white noise perturbation $\rho_{n1}$ is introduced into the neutral density profile to initiate RTI development.  This perturbation has the form
\begin{equation} 
\rho_{\rm n1} = 10^{-2}\rho_{\rm n0} \times r_{\rm x} \exp{\left[-(2z/L_0)^2\right]},
\end{equation}
where $r_{\rm x}$ is a random number in the interval \mbox{[-1,1]} for each of the grid points.

For all simulations, the numerical box of 2~Mm $\times$ 8~Mm is uniformly discretized with 2048 $\times$ 8192 grid points, $\Delta x = \Delta z = 0.977$~km, which is the same resolution as for the simulation discussed in \cite{rti3} and four times higher resolution than in \cite{rti1,rti2}.  The boundary conditions are periodic in the $x$-direction.  In the $z$-direction, we use anti-symmetric boundary conditions for the vertical velocity of charges and neutrals and symmetric for the rest of the variables, with the $\hat{z}$-boundaries located sufficiently far from the prominence thread to minimally impact the dynamics of the simulations.

\subsection{Global RTI evolution}

\begin{figure}
\centering
\includegraphics[width=0.88\textwidth]{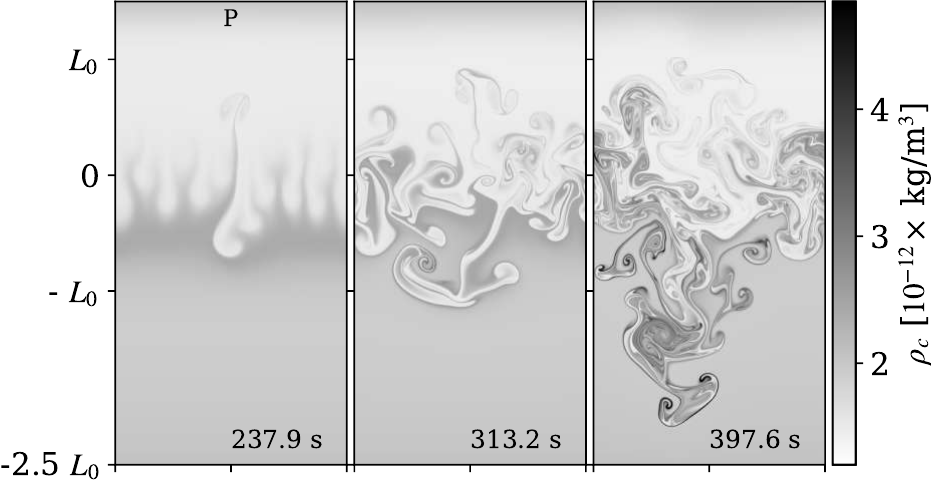}
\includegraphics[width=0.88\textwidth]{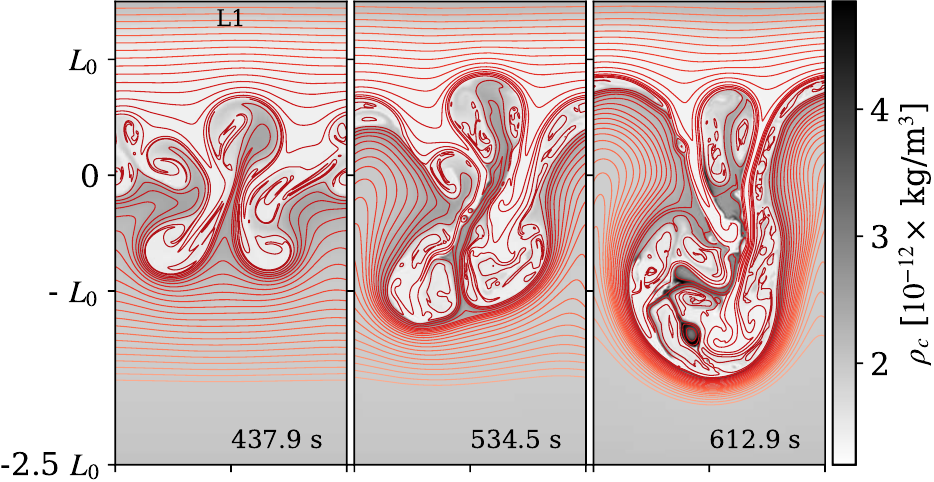}
\includegraphics[width=0.88\textwidth]{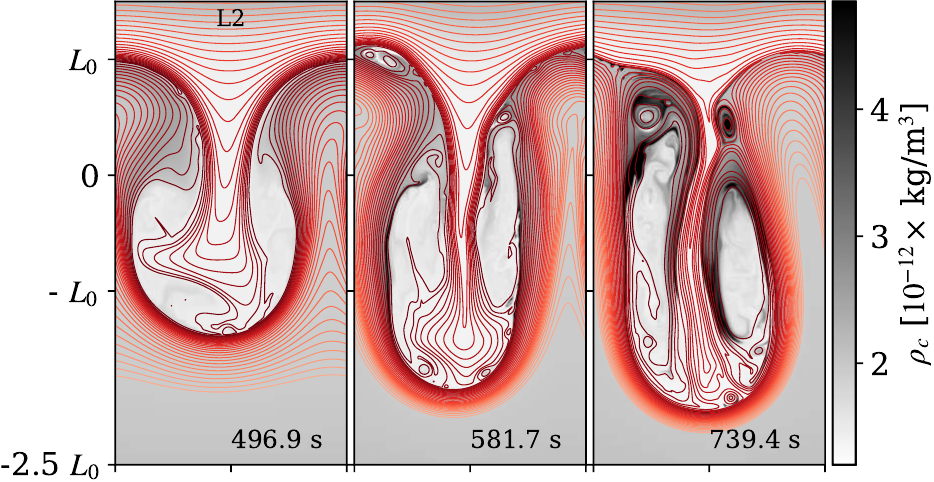}
\caption{Time evolution of densities of charges (grey scale) and in-plane magnetic field structure (red lines) for the P (top), L1 (middle) and L2 (bottom) simulations. The time of the snapshots is indicated at each panel; notice that the snapshots from each series are not co-temporal. For clarity, only a part of the vertical extend of the simulation domain from -2.5$L_0$ to 1.5 $L_0$ is shown.}
\label{fig:RTIsnaps}
\end{figure}

The qualitative differences among the simulations are shown in Figure~\ref{fig:RTIsnaps}. The representative snapshots are from three simulations each initialized with one of the three magnetic field configurations: P (top row), L1 (middle row), and L2 (bottom row).  The evolution of the mass density of charges is shown in a grey color scale with the in-plane magnetic field structure over-plotted for L1 and L2, while no in-plane magnetic fields are present in the P simulation.  Only the central part, $z\in [-2.5 L_0, 1.5 L_0]$, of the full vertical extent of the simulation domain is shown for clarity of presentation.  For L1 and L2 snapshots, evolution of the same field lines is tracked across the snapshots with the field-line density corresponding to the strength of the in-plane magnetic field. The plotted field lines are equidistant in the $A_{\rm y}$ vector potential space with $\mathbf{B}_{\rm xz} = -\nabla\times(A_{\rm y}\hat{y})$, with the same number of field lines plotted above and below the $|\mathbf{B}_{\rm xz}|=0$ contour. For P and L1 simulations, the right-most panels at $t=397.6$~sec and $t=612.9$~sec, respectively, represent the latest available snapshots before the simulations were terminated due to the development of under-resolved scales in the system\footnote{We note that the $t=397.6$~sec and $t=612.9$~sec snapshots for the P and L1 simulations, respectively, do not represent the final time-steps in the simulations. Rather, these represent the last of the periodic data outputs generated by the code while the simulations were deemed to be sufficiently converged to continue running.}; the L2 simulation, on the other hand, was terminated at an arbitrary time after the saturation of the instability.

The differences in the RTI evolution across the three cases are apparent.  In the absence of the magnetic field in the mixing plane, the P simulation shows fine-scale mixing between the hot and cold plasma regions with secondary and tertiary RTI and KHI structures freely developing.  We note that the regions of low charge mass density at $t=237.9$~sec and $t=313.2$~sec correspond to the cold dominantly neutral prominence material falling through the PCTR. However, by $t=397.6$~sec, these same falling down plasma elements have evolved to gain substantially more charged mass density than even the surrounding hotter medium.  The primary reason for this evolution is the heating and subsequent ionization at the edges of the cold dense prominence structures.  As a result of the extensive mixing, the effective surface area of the hot-cold interface becomes large, while the sharp gradients across the interface are only contained by collisional dissipative processes.

The L1 simulation has already been extensively discussed in \cite{rti3}, so here we only re-emphasize the rich in-plane magnetic field structure generated within the PCTR as a result of the RTI development.  We again observe formation of some structures with charge mass density above the background coronal medium, but in this case these are due primarily to magnetic field reconnection processes described in \cite{rti3}.  There is substantially less direct mixing between the cold prominence material and the hotter coronal background plasma in the L1 than in the P simulation.

The L2 simulation, with the magnetic shear scale equal to the prominence density gradient scale that drives the RTI, shows development of a single primary mode that saturates with limited magnetic field sub-structure and even less mixing between the cold and hot material than in the L1 simulation.  The interface between the prominence and the coronal material, contained by the highly compressed magnetic field at the bottom, shows formation of some charge mass density and magnetic field substructure via magnetic reconnection near the top of the plume by $t=739.4$~sec.  Nevertheless, most of the volume of the resulting PCTR is devoid of either magnetic of charge density substructure.

\begin{figure}
\centering
\includegraphics[width=\textwidth]{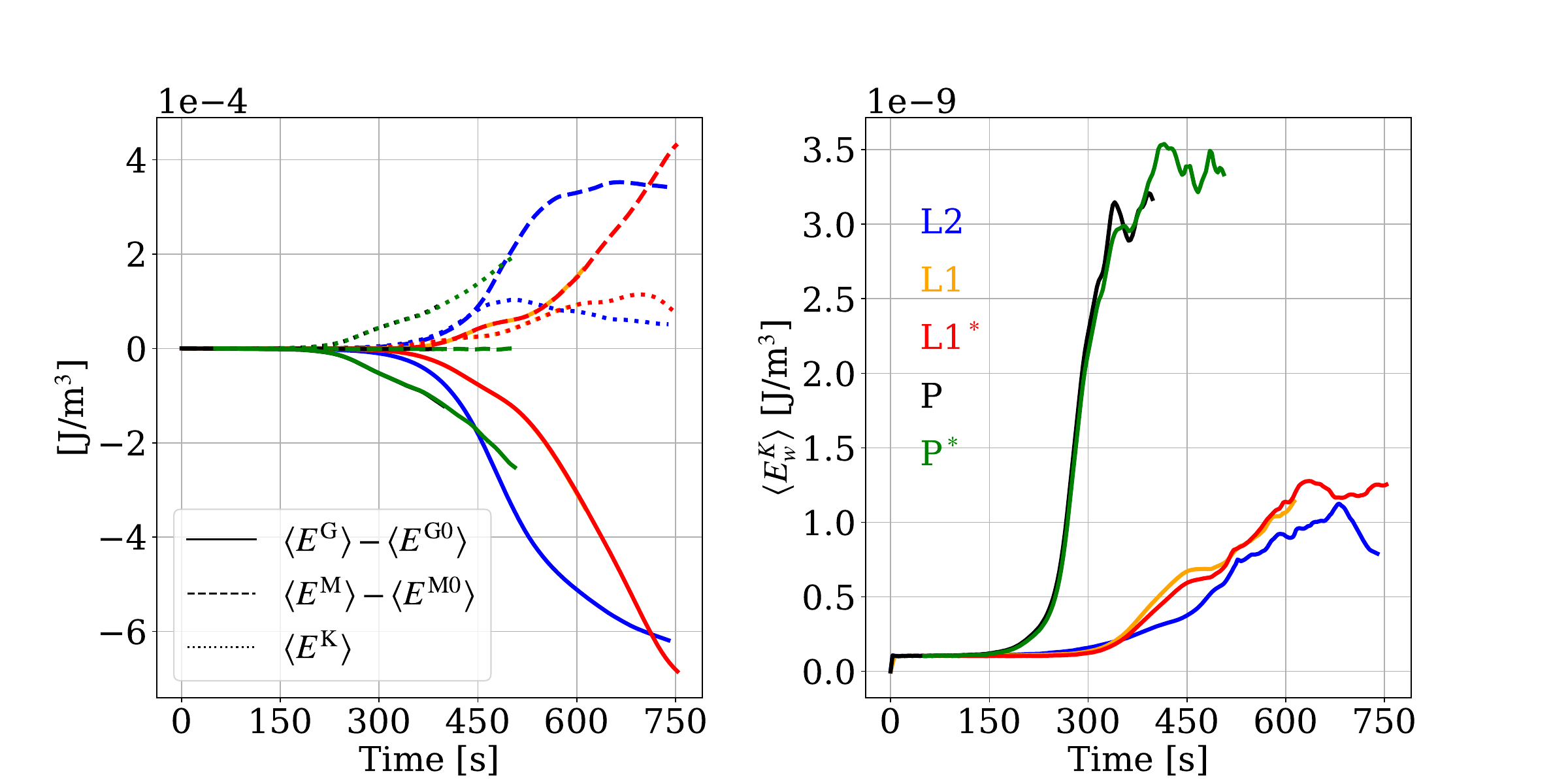}
\caption{Time evolution of domain-averaged energy components for five simulations P/P*, L1/L1*, and L2. Left: Change of the potential gravitational energy of the plasma, $E^G$, relative to the initial condition (solid lines); change of the magnetic energy in the domain, $E^M$, relative to the initial condition (dashed lines); total kinetic energy of the plasma in the domain, $E^K$ (dotted lines). Right: Kinetic energy contained in the relative drift between neutrals and charges, $E^K_{\rm w}$.
}
\label{fig:evTime}
\end{figure}

The temporal evolution of the RTI non-linear development for the three initial magnetic field configurations is analyzed from the perspective of energy transfer from the gravitational, $E^G$, to the magnetic, $E^M$, and the kinetic, $E^K$, energy components in Figure~\ref{fig:evTime}.  To be able to follow the dynamics of RTI evolution for the P and L1 initial configurations further in time, we conducted two additional simulations, denoted as P$^*$ and L1$^*$, where the ionization and recombination reactions in the governing equations were turned off after $t=50$~sec.  Doing so eliminated the formation of high charge mass density interfacial layers described above and allowed the simulations to proceed substantially further, with the P$^*$ simulation running for an additional $\approx 100$~sec, and the L1$^*$ simulation proceeding through to the non-linear RTI saturation.  We note that eliminating the ionization/recombination reactions in these simulations changes the evolution of potential observables, such as the relative amount of ionized material in the high density structures.  However, the largely overlapping energy component traces for P and P$^*$, and L and L1$^*$, shown in Fig.~\ref{fig:evTime} indicate that the global dynamics of the RTI is not significantly impacted by the non-equilibrium ionization processes in our simulations.

We showed in \cite{rti1} that the linear RTI growth rate of the fastest growing mode is highest for the P configuration, followed by L1, and then L2.  The timing of the onset of the non-linear transfer of the gravitational energy initially stored in the prominence material to the kinetic and, in L1 and L2 cases, magnetic energy within the simulation domain is shown in the left panel of Fig.~\ref{fig:evTime}.  While the P (and P$^*$) simulation is the first to show non-linear energy transfer, the L2 configuration appears to allow for faster gravitational energy release than L1.  Further, the rate of L2 gravitational energy release is such that it rapidly overtakes the same for the P simulation. That is, the nonlinear energy release is the fastest in the configuration with the slowest linear instability growth rate!  While appearing to be counter-intuitive, similar behavior has been previously observed in RTI simulations by  \cite{Stone2007b}, where they credit the magnetic field with inhibiting secondary instabilities and reducing the mixing to cause the rate of non-linear growth of the primary bubbles and fingers to increase.  It is consistent with what we observe by comparing P and L2 snapshots in Fig.~\ref{fig:RTIsnaps}.

The bulk of the released gravitational energy in both L1 and L2 configurations is deposited into the magnetic field energy, which is what we previously showed for L1 in \cite{rti3}.  The L1$^*$ simulation evolved through the saturation of the RTI kinetic energy indicates that a broader magnetic field shearing layer allows the prominence material to fall even further, and for more magnetic field energy to be built up before the instability saturates.

On the other hand, the RTI development in the P configuration deposits the majority of the released gravitational energy into the kinetic energy of the plasma.  However, as the instability develops, this kinetic energy is stored increasingly in the turbulent eddy flow of the mixing layer rather than the bulk flow of the dropping prominence material.  While the gain in the total kinetic energy contained in the flow accelerates with time, the rate at which the gravitational energy is decreasing is nearly constant, meaning that the kinetic energy associated with the motion of the center of mass of the prominence material also stays nearly constant in time.

The right panel of Fig.~\ref{fig:evTime} shows the amount of the kinetic energy contained in the relative drift between neutrals and charges, $\mathbf{w} \equiv \mathbf{u_n} - \mathbf{u_c}$ for the same simulations.  We define the kinetic energy in the relative drift between neutrals and charges, $E^K_{w}$, as the energy difference between the sum of the individual kinetic energies and the kinetic energy both fluids would have if co-moving at the center of mass velocity $\mathbf{u}\equiv \frac{\rho_n \mathbf{u_n} + \rho_c \mathbf{u_c}}{\rho_n + \rho_c}$:
\begin{equation}
    E^K_{\rm w} \equiv (E^K_n + E^K_c) - E^K_{u} = \frac{1}{2}\frac{\rho_n \rho_c (\mathbf{u_n} - \mathbf{u_c})^2}{\rho_n + \rho_c}
\end{equation}
We observe that the magnitude of $E^K_w$ is in all cases five orders of magnitude smaller than the total kinetic energy of the flows, i.e. the flows are dominantly well-coupled.  However, we also observe that the fraction of the kinetic energy contained in the charges-neutral drift in the P configuration simulations is up to twice that for the L1 and L2 configurations.  This indicates that the flow decoupling during RTI development is likely greatest in the plane normal to the direction of the magnetic field, which is what we explore further below.

\section{Ion-neutral drift signatures}

As discussed in \cite{rti1,rti2,rti3}, the dynamical decoupling between charges and neutrals is particularly pronounced at the interfaces between the prominence material and the coronal plasma. As the instability progresses and nonlinear flows expand across the domain, the surrounding transition region lengthens, notably in the P case where the flow develops in the plane normal to the magnetic field.

\begin{figure}
\centering
\includegraphics[width=1.0\textwidth]{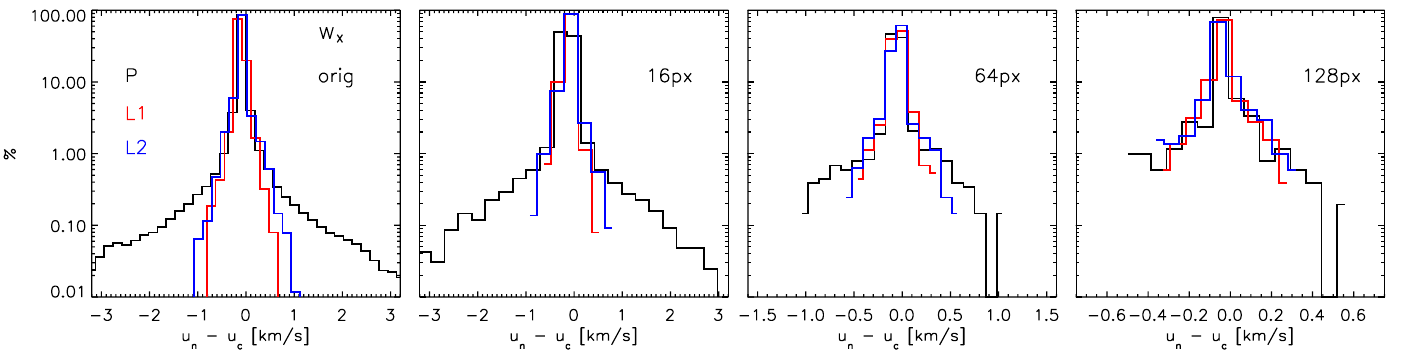}
\includegraphics[width=1.0\textwidth]{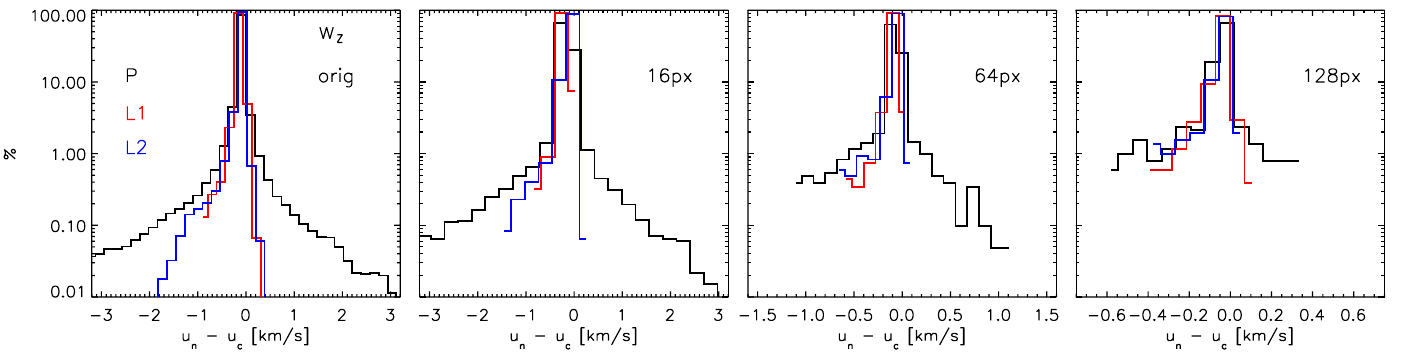}
\caption{Histograms of the drift velocities, $\mathbf w=\mathbf{u}_n-\mathbf{u}_c$, for the original simulation resolution (left), and for degraded resolutions (from the second panel to the right), computed as simple algebraic averaging, $\langle\mathbf{w}\rangle=\sum{\mathbf{w}_{i,j}}/px^2$, where $px$ represents the linear pixel size as indicated in each panel. The histograms are computed for the last available simulation time for each magnetic configuration, as shown in  Fig.~\ref{fig:RTIsnaps}: black lines are for the P case at $t=397.6$~sec, red are for the L1 case at $t=612.9$~sec, and blue are for the L2 case at $t=739.4$~sec.
Upper panels: horizontal component, $\langle\rm w_x\rangle$; bottom panels: vertical component, $\langle\rm w_z\rangle$.}
\label{fig:hist_dec}
\end{figure}

Figure~\ref{fig:hist_dec} illustrates histograms of the decoupling velocity, $\mathbf{w}=\mathbf{u}_n - \mathbf{u}_c$, depicting horizontal velocity (upper panels) and vertical velocity (bottom panels) at four different resolutions for the three magnetic configurations: P (black), L1 (red), and L2 (blue). The leftmost panels represent the original resolution of the simulations, where the pixel size is approximately $1\times 1$ km$^2$. It's evident that both horizontal and vertical decoupling are considerably more pronounced in the P case, showing near symmetry with extreme values reaching $\pm 3$~km/s and higher. As discussed below, this decoupling is primarily due to the uncompensated neutral pressure gradient.

In the L1 and L2 cases, where magnetic forces are also influential but the structures formed are larger, the decoupling velocities exhibit lower amplitudes, not exceeding $\pm 2$~km/s. It's worth noting that the L1 and L2 vertical drift histograms are strongly asymmetric, with the negative velocities dominating, while the P case shows a much weaker vertical asymmetry. These negative drift velocities indicate the neutrals falling slightly faster than the ions; and the weak vertical asymmetry of the ion-neutral drift histogram for the P configuration confirms the dominance of the turbulent eddy flow over the bulk downward motion in the plane normal to the magnetic field.  This observation also aligns with our earlier findings reported in \cite{rti2} from simulations with four times lower resolution (note the considerably smaller drifts obtained in that case). 

Going form left to right in Fig.~\ref{fig:hist_dec}, histogram panels display the same velocities but with progressively coarsened pixels, with pixel sizes of $16\times 16$~km$^2$, $64\times 64$~km$^2$, and $128\times 128$~km$^2$. We note that the range of the measured values for $\rm w_x$ and $\rm w_z$ on the horizontal axes progressively contracts from left to right. The velocities are averaged using simple algebraic averaging, calculated as $\langle{\rm w}\rangle=\sum{{\rm w}_{i,j}}/px^2$, where $px$ represents the linear pixel size. We observe that at the $16\times 16$~km$^2$ resolution the original distribution remains adequately preserved, indicating that spatially our simulations are sufficiently resolved. However, at the coarser resolutions, the histograms' tails are significantly truncated, eliminating the largest drifts. This effect arises because the most substantial drifts concentrate in narrow layers, which, upon averaging, become "washed-out." At the coarsest resolution, the expected detectable range of drifts is approximately $\pm 0.4$~km/s, and the asymmetry in the vertical velocity histogram remains discernible. This resolution aligns with what might be achievable by instruments like DKIST and possibly SST.

\begin{figure}
\centering
\includegraphics[width=0.8\textwidth]{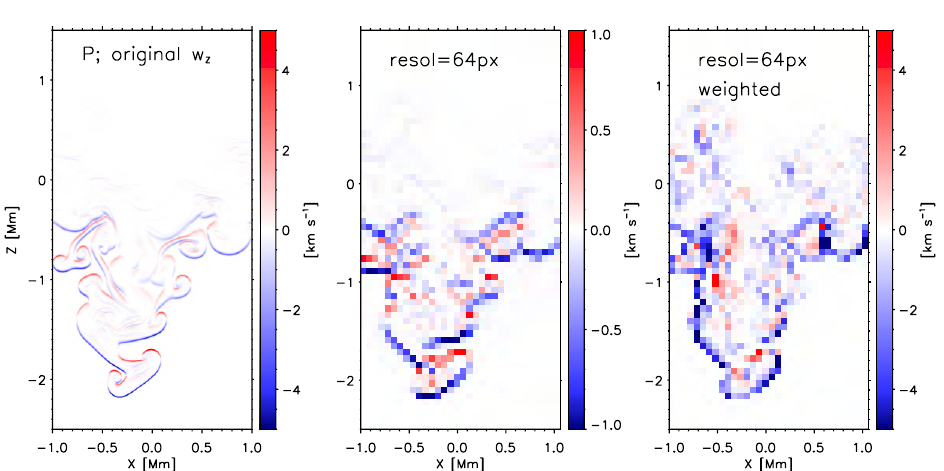}
\includegraphics[width=0.8\textwidth]{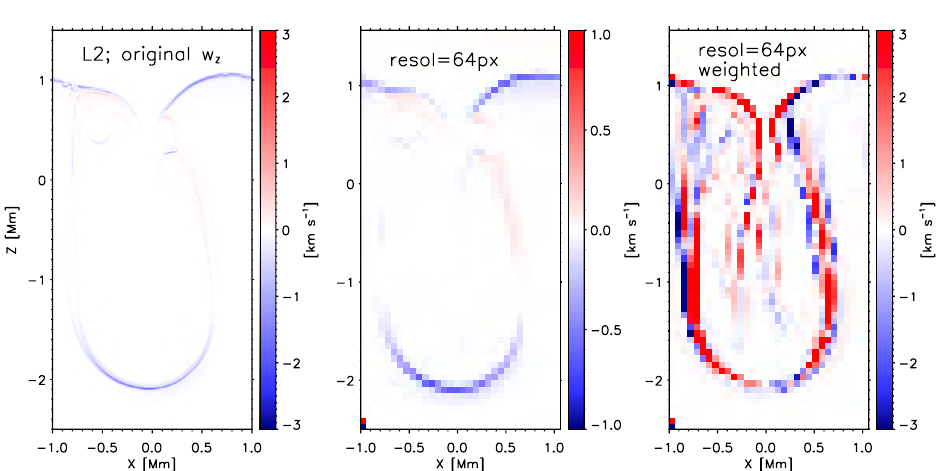}
\caption{Snapshots of the drift velocities in the P case (top) and L2 case (bottom) for the latest available simulation times corresponding to the right-most panels in Fig.~\ref{fig:RTIsnaps}. For each case, the panels from left to right show the original snapshot (left) and two coarsened snapshots with the resolution of 64$^2$ px by averaging the drift velocities either directly, $\langle{\rm w}\rangle =\sum{(u_n-u_c)}/64^2$ (middle), or with weighted averaging,  $\langle{\rm w}\rangle_\rho = \sum{(u_n\rho_n)}/\sum{\rho_n} - \sum{(u_c\rho_c)}/\sum{\rho_c}$ (right).}
\label{fig:decoupling}
\end{figure}

\begin{figure}
\centering
\includegraphics[width=0.4\textwidth]{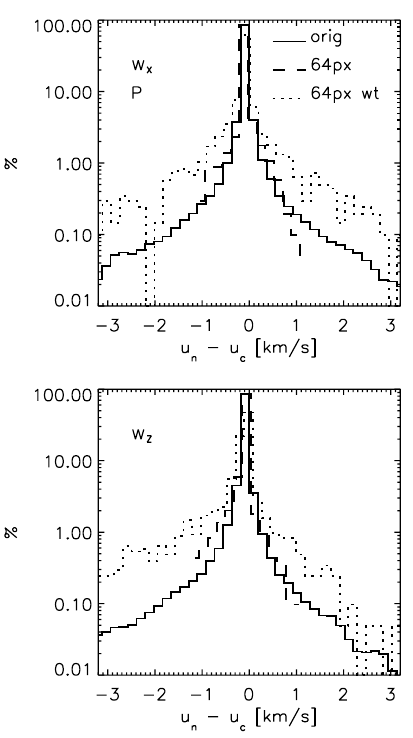}
\includegraphics[width=0.4\textwidth]{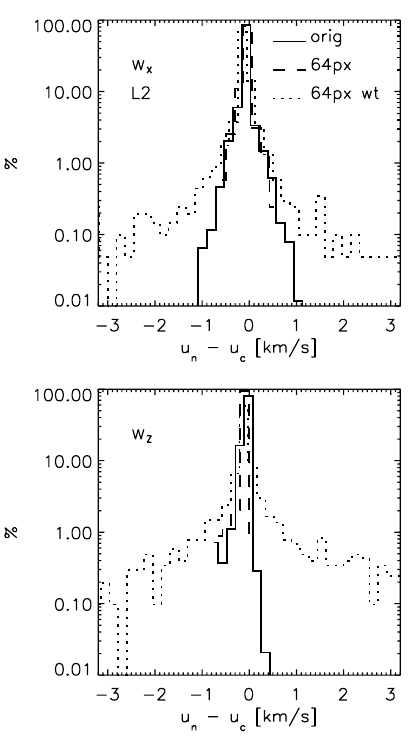}
\caption{Histograms of the horizontal (top) and vertical (bottom) drift velocities for the P case (left) and L2 case (right) for the latest available simulation times corresponding to the right-most panels in Fig.~\ref{fig:RTIsnaps}. Solid line is for the original resolution, dashed and dotted are for the 64$^2$ px coarsened resolution. The dashed lines are for the simple algebraic averaging, while the dotted ones are for the density-weighted averaging. }
\label{fig:hist_dec_rho}
\end{figure}

Figures~\ref{fig:decoupling} and \ref{fig:hist_dec_rho} illustrate the comparison between different approaches for averaging drift velocities. In observational studies \citep{2016Khomenko, 2017Anan, Wiehr2019, Wiehr2021, Zapior2022, GonzalezManrique2023}, velocities are derived from spectral lines of neutral and ionized elements. Due to finite spatial resolution, the spectral signal in an observational pixel, emanating from an optically thin plasma, is a blend of signals along the line of sight and across the horizontal surface of the pixel. This signal is weighted by the corresponding spectral line amplitudes.  In an attempt to mimic this observational procedure, we computed an alternative proxy for mean drift velocity, $\langle \mathbf{w}\rangle_{\rho} \equiv \sum{(\mathbf{u}_n \rho_n)}/\sum{\rho_n} - \sum{(\mathbf{u}_c \rho_c)}/\sum{\rho_c}$, with the sum taken over all computational grid points within a single coarsened pixel. This proxy incorporates a weighting factor proportional to the mass density of the corresponding component (neutral or ionized), which might resemble the amplitude of a typical spectral line.

Figure~\ref{fig:decoupling} compares snapshots of ${\rm w}_z$ at the original resolution (left), with the simple algebraic averaging (middle), and with the density-weighted averaging (right) for the P case (top) and the L2 case (bottom). (We note that the color bar values in the middle panel are reduced compared to the left and right panels.) The figure reveals that in both cases the largest decoupling occurs at the interface layer surrounding the PCTR. For both, the algebraic averaging shown in the middle panels reduces the amplitude of ${\rm w}_z$ without altering its spatial distribution; it essentially blurs the image.  However, the density-weighted averaging shown in the right panels significantly distorts information from the original snapshots. This method amplifies the resulting ${\rm w}_z$ and occasionally even changes the sign of the locally computed drift. 

There is an intuitively simple reason for the potential distortion of the drift measurement when using density-weighted averaging.  In a larger pixel, the locations with the largest neutral and charge velocities and densities are not necessarily spatially overlapping. Algebraic averaging consistently subtracts $\mathbf{u}_n$ and $\mathbf{u}_c$ at the same location. By contrast, the weighted averaging enhances the impact of the values of $\mathbf{u}_n$ and $\mathbf{u}_c$ at the locations with the high $\rho_n$ and $\rho_c$, respectively, which might not coincide for neutrals and charges. This discrepancy becomes evident at the narrow hot-cold plasma interfaces where the dominantly ionized plasma on the one side meets the dominantly neutral fluid on the other side. A large pixel encompassing this border would average $\mathbf{u}_n$ from the cold side of the interface and $\mathbf{u}_c$ from the hot side of the interface, resulting in non-co-spatial averages. This leads to the artificial enhancement of $\langle{\rm w}_z\rangle_\rho$ seen in the right panels of Fig.~\ref{fig:decoupling}. 

Figure~\ref{fig:hist_dec_rho} demonstrates the same effect via the histogram analysis, showing that a density-weighted average $\langle{...}\rangle_\rho$ may not accurately represent the original distribution. The dotted lines in Fig.~\ref{fig:hist_dec_rho} show that the histograms of both $\langle{\rm w}_x\rangle_\rho$ and $\langle{\rm w}_z\rangle_\rho$ for both P and L2 cases develop artificial wings toward larger values not present in the original data (solid lines) or the algebraic averages (dashed lines). The implied drifts in the L2 case are particularly affected, resulting in very similar horizontal and vertical drift distributions, as well as eliminating the qualitative difference between the P and L2 cases in the original data. This simple exercise suggests that observed drift magnitudes need careful interpretation. While observations like \cite{2016Khomenko, 2017Anan, Wiehr2019, Wiehr2021, Zapior2022, GonzalezManrique2023} may indeed suggest plasma-neutral decoupling, the magnitude of this decoupling might be artificially inflated by the observational procedure.

\section{Heating mechanisms}
\label{sec:heating}

We now return to the analysis of the energy transport during RTI development in the presented simulations with a focus on the heating mechanisms and the localization of where the heating takes place.

The potentially important irreversible heating mechanisms in our two-fluid model are the viscous heating for neutrals, $Q^{\rm VISC}_{\rm n}$, and charges, $Q^{\rm VISC}_{\rm c}$, the frictional heating between the neutral and charged fluids, $Q^{\rm FRIC}$, and the effective heating resulting from the numerical dissipation of magnetic fields, $Q^{\rm MAG}_{\rm num}$, and flows, $Q^{\rm VISC}_{\rm num}$, implemented in \mancha as a 6th order spatial filtering of the evolved variables at some prescribed temporal cadence \citep{Popescu+etal2018, rti3}.  In addition, adiabatic heating and cooling, $Q^{\rm COMP}$, as well as the thermal conduction of the neutrals, $K_n \nabla T_n$, and the thermal exchange between neutrals and charges can play important roles in determining the temperature evolution and transport of the internal energy in the system.  

We note that the two-fluid model used in this work did not include the thermal conduction for the charges, which is reasonably justified for the considered magnetic configurations with at most $\approx 1\%$ of the magnetic field projected into the simulation plane.  We also note that for the considered plasma parameters, the local thermal exchange between neutrals and charges is a very fast process everywhere in the computational domain, such that the temperatures of the two fluids locally equilibrate faster than any other thermal transport or heating process in the system.

The functional forms in the \mancha two-fluid equations for the heating and internal energy transport terms referenced above are as follows: \\
\noindent
The viscous heating terms for the neutral and charged fluids are calculated as:
\begin{equation}
\label{eq:qvisc}
Q^{\rm VISC}_\alpha = \mathbf{\hat{\tau}}_\alpha:\nabla\mathbf{u}_\alpha \, 
\text{, with viscous tensor components } \tau_{\alpha ij} \equiv \xi_\alpha \left(\frac{\partial u_{\alpha i}}{\partial x_j} + \frac{\partial u_{\alpha j}}{\partial x_i} \right) \,,
\end{equation}
where the neutrals and charges viscosity coefficients are:
\begin{equation}
\label{eq:visc_coef}
  \xi_n =  \frac{ \sqrt{k_B T_n \pi m_H}}{2 \Sigma_{\rm nn} } \text{  and  }
\xi_c = \frac{\sqrt{k_B T_c \pi m_H}}{2 \Sigma_{\rm ii}} .
\end{equation}
The frictional heating between neutrals and charges is calculated as
\begin{equation}
    Q^{\rm FRIC}=\alpha\rho_n\rho_c(\mathbf{u}_c-\mathbf{u}_n)^2
\end{equation}
with the collisional parameter $\alpha$ given by Eqs.~(5-9) in \cite{rti3}. \\
\noindent
The effective heating due to the numerical dissipation of magnetic fields can be approximated as
\begin{equation}
\label{eq:qjoule_num}
Q^{\rm MAG}_{\rm num}=
\frac{\eta_6}{\mu_0} \sum_{j=x,y,z}{\left|\nabla (\nabla^2 B_{j})\right|^2}
\end{equation}
and the corresponding heating due to the numerical dissipation of the flows can be approximated as
\begin{equation}
   Q^{\rm VISC}_{\rm num} = \eta_6 \sum_{j=x,y,z}\left\{ \frac{1}{\rho_n}\left|\nabla[\nabla^2(\rho_n u_{nj})]\right|^2 + \frac{1}{\rho_c}\left|\nabla[\nabla^2(\rho_c u_{cj})]\right|^2  \right\},
\end{equation}
with $\eta_6=4.7\times10^{18}$~m$^6$/s, as derived in \cite{rti3}. \\
\noindent 
The total compressional heating/cooling combined for the two fluids is calculated as
\begin{equation}
\label{eq:compheat}
    Q^{\rm COMP}= -(p_n \nabla \cdot \mathbf{u}_n + p_c \nabla \cdot \mathbf{u}_c)
\end{equation}
and the neutral thermal conductivity coefficient is calculated as:
\begin{equation}
\label{eq:heatcond}
K_n = \frac{2 k_B  \sqrt{\frac{\pi k_B  T_n}{m_H}}}{\Sigma_{\rm nn}}\,.
\end{equation}
The collisional cross sections used in Eq.~\ref{eq:visc_coef} and Eq.~\ref{eq:heatcond} are:
\begin{equation}
    \Sigma_{\rm nn}=7.73 \times 10^{-19} {\rm m}^2 \text{  and   } 
    \Sigma_{\rm ii}=\frac{20\pi\sqrt{2}}{3} 
\left( \frac{e^2}{4 \pi \epsilon_0 k_B T_c} \right)^2 {\rm m}^2 .
\end{equation}
We note that incorrect expressions for these cross sections, and for $K_n$, were stated in the previous paper \citep{rti3}; in fact, simulation results presented in all three prior papers on RTI modeling, \cite{rti1, rti2, rti3}, used the $\Sigma_{\rm nn}$, $\Sigma_{\rm ii}$, and $K_n$ expressions stated above. 

\begin{figure}
\centering
\includegraphics[width=\textwidth]{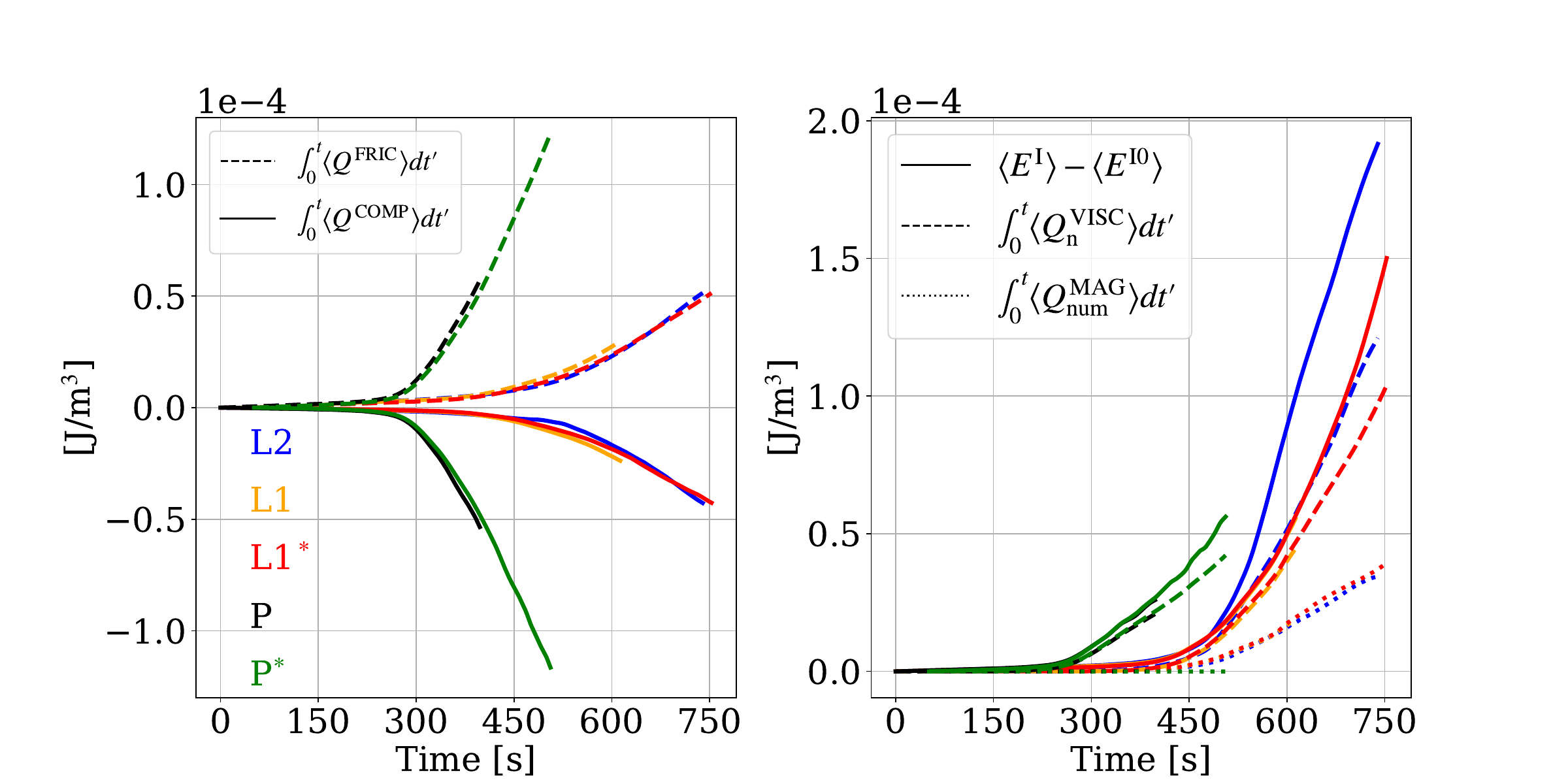}
\caption{Time evolution of domain-averaged plasma heating terms for five simulations P/P*, L1/L1*, and L2. Left: Change in the internal energy due to the frictional heating, $Q^{\rm FRIC}$ (dashed lines), and the compressional heating/cooling, $Q^{\rm COMP}$ (solid lines). Right: Total change in the internal energy, $E^I$, relative to the initial condition (solid lines); change in the internal energy due to viscous heating of neutrals $Q^{\rm VISC}_{\rm n}$ (dashed lines), and due to numerical dissipation of magnetic field $Q^{\rm MAG}_{\rm num}$ (dotted lines). }
\label{fig:QvTime}
\end{figure}

To explore the roles of the different heating and thermal transport mechanisms in the RTI simulations, we first analyze the temporal evolution of domain-integrated quantities for the three magnetic field configurations.  Figure~\ref{fig:QvTime} details the time-accumulated impact of domain-averaged plasma heating terms for all five simulations, P/P*, L1/L1*, and L2, described in Fig.~\ref{fig:evTime}.  It could have been expected that the frictional heating between charges and neutrals, $Q^{\rm FRIC}$, may significantly contribute to the overall heating of the plasma.  Instead, we find in the left panel of Fig.~\ref{fig:QvTime} that the time-integrated frictional heating is at all times almost exactly compensated by the time-integrated compressional cooling, $Q^{\rm COMP}$, for each of the five simulations.  We also find that the compressional cooling is primarily due to the evolution of the neutral fluid, with a negligible contribution of the compressional term for the charges (not shown).  

It is notable that both $Q^{\rm FRIC}$ and $Q^{\rm COMP}$ have the largest domain-averaged magnitude for the P magnetic configuration, while the simulations in the L1 and L2 configurations appear to have nearly identical domain-averaged time evolution for these quantities.  The former is not surprising as we have already demonstrated that RTI development in the plane normal to the magnetic field allows for strongest ion-neutral drifts. The latter, however, is less intuitive with markedly different energy time traces for the L1 and L2 configuration simulations shown in the left panel of Fig.~\ref{fig:evTime}; though we observe that the $E^K_{\rm w}$ traces for L1 and L2 in the right panel of Fig.~\ref{fig:evTime} also show significant similarity, with the energy source for the frictional heating ultimately being the differential forcing of the flows of the two coupled fluids.

The right panel of Fig.~\ref{fig:QvTime} shows the accumulation of the internal thermal energy from all sources, and the $Q^{\rm VISC}_{\rm n}$ and $Q^{\rm MAG}_{\rm num}$ contributions to the heating.  It is apparent that in all cases the net heating which contributes to the increase of the internal energy comes dominantly from neutral viscous heating, and to a smaller extent from the numerical magnetic dissipation for the L1 and L2 magnetic configurations which again track each other.  Taken together, $Q^{\rm VISC}_{\rm n}$ and $Q^{\rm MAG}_{\rm num}$ account for the vast majority of the measured increase in the internal thermal energy for all three magnetic configurations.  Nevertheless, we once again note a marked difference between the P configuration on one hand, and the L1 and L2 configurations on the other.  While RTI evolution in the plane normal to the magnetic field produces more than twice frictional than viscous heating, the viscous heating dominates over the frictional heating by approximately the same factor for the L1 and L2 configurations.  Even more surprisingly, while the P configuration produces more than twice as much kinetic energy in the turbulent eddies that is available to be viscously dissipated (see discussion of Fig.~\ref{fig:evTime}), the viscous dissipation in the L1 and L2 cases leads to more than twice as much actual viscous heating than measured in the P case.

\begin{figure}
\centering
\includegraphics[width=1.0\textwidth]{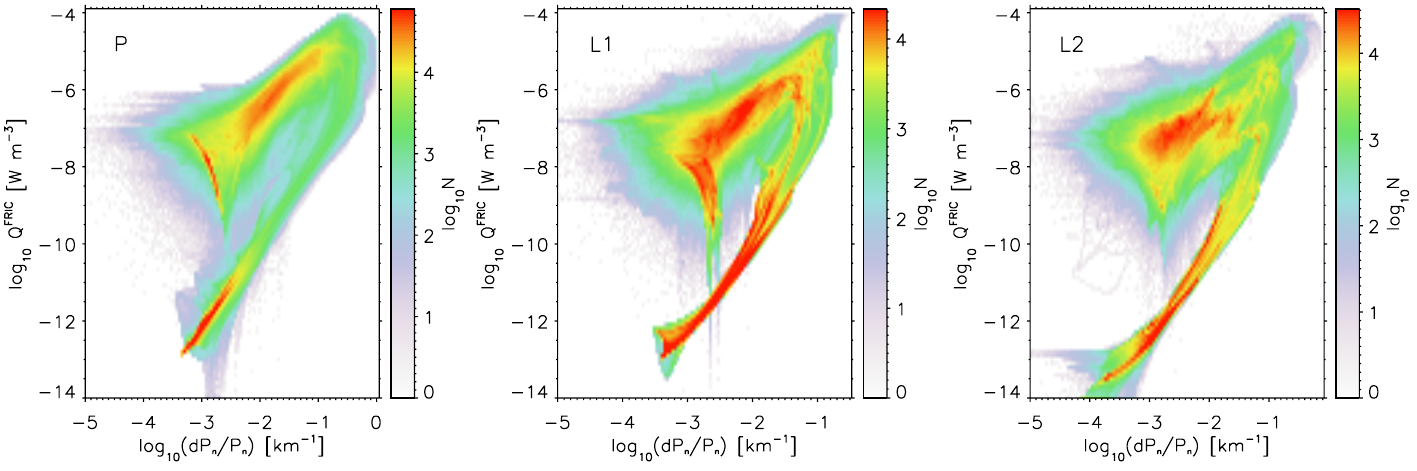}
\caption{Log-log correlations between the frictional heating, $Q^{\rm FRIC}$, and the inverse neutral pressure gradient scale, $H_{p_n}^{-1} \equiv |\nabla p_n|/p_n$ for the P (left), L1 (middle) and L2 (right) cases as evaluated at the latest available simulations times corresponding to the right-most panels in Fig.~\ref{fig:RTIsnaps}. The color coding indicates the number of points in the distribution on the logarithmic scale. }
\label{fig:corr}
\end{figure}

\begin{figure}
\centering
\includegraphics[width=1.0\textwidth]{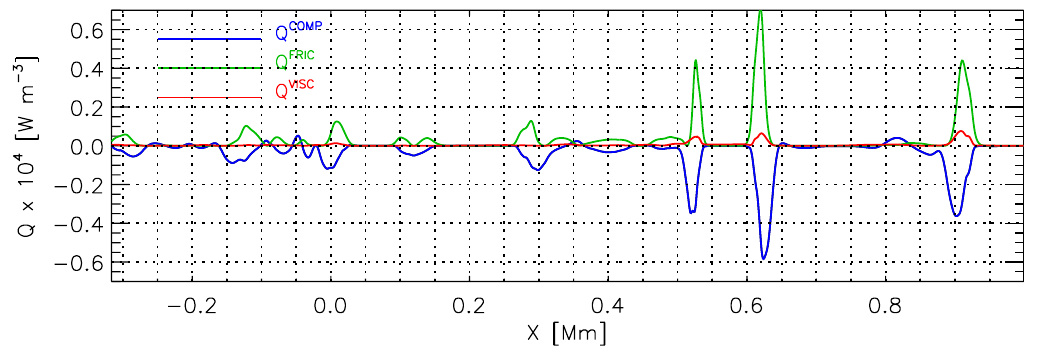}
\caption{Illustration of the heating terms in the P simulation using a 1D cut at $z=-0.68$~Mm at $t=397.6$~sec. $Q^{\rm COMP}$ is shown in blue, $Q^{\rm FRIC}$ in green, and $Q^{\rm VISC}$ in red.}
\label{fig:qcomp_qfric_qvisc}
\end{figure}

To better understand the local sources of heating and cooling, we study these in more detail.  Figure~\ref{fig:corr} illustrates the correlations between a characteristic local forcing scale determined by the neutral pressure gradient, $H_{\rm p_n}^{-1}=|\nabla p_n|/p_n$, and frictional heating, $Q^{\rm FRIC}$, for each of the magnetic configurations computed at the latest available simulation times corresponding to the right-most panels in Fig.~\ref{fig:RTIsnaps}. There is a significant correlation between the two quantities, showing the amplification of frictional heating in regions where the neutral pressure gradients are larger and neutrals experience compressional forces over shorter scales.  As discussed in \cite{rti2,rti3}, it is the neutral pressure gradients that are the dominant drivers of ion-neutral drifts, and therefore also of the resulting friction and frictional heating, in our RTI simulations.  
Combined with the new finding of the apparent persistent balance in the overall compressional cooling and the frictional heating shown in Fig~\ref{fig:QvTime}, we can reasonably conclude that the expansion of the cold dense plasma at the sharp hot-cold interfaces embedded in the PCTR may lead to the simultaneous compressional cooling and frictional heating of this interface region.  We illustrate, and confirm, this hypothesis in Figure~\ref{fig:qcomp_qfric_qvisc} showing $Q^{\rm COMP}$, $Q^{\rm FRIC}$, and $Q^{\rm VISC}$ along a partial cut through the simulation domain for the P magnetic configuration simulation at $t=397.6$~sec.  The cut crosses several cold dense plasma structures (see the top-right panel of Fig.~\ref{fig:RTIsnaps} for reference) and demonstrates that the frictional heating and compressional cooling indeed act in equal and opposite directions with slight displacement relative to, but generally in the same locations as each other.  The viscous heating is much weaker, but similarly co-located.

Another notable feature in all three of Fig.~\ref{fig:corr} panels for the P, L1, and L2 configurations is the presence of two distinct distributions: a broader one for larger values of $Q^{\rm FRIC}$ and a narrower one beneath it. The narrower distribution originates from the initial background characterized by large-scale pressure gradients established in the prominence thread equilibrium \citep{rti1}. As the initial equilibrium undergoes mixing, a secondary distribution emerges, reflective of the nonlinear stage of the instability. This new distribution features notably larger values of $Q^{\rm FRIC}$. Similar qualitative dependencies are observed in the P, L1, and L2 cases. However, it's important to note that the distribution maximum (indicated by the location of the red points) gradually shifts towards lower $Q^{\rm FRIC}$ values from P to L1 and L2 cases. This shift indicates not only lower values of $Q^{\rm FRIC}$ in the L1 and L2 simulations but also corresponds to more large-scale structures.

\begin{figure}
\centering
\includegraphics[width=1.0\textwidth]{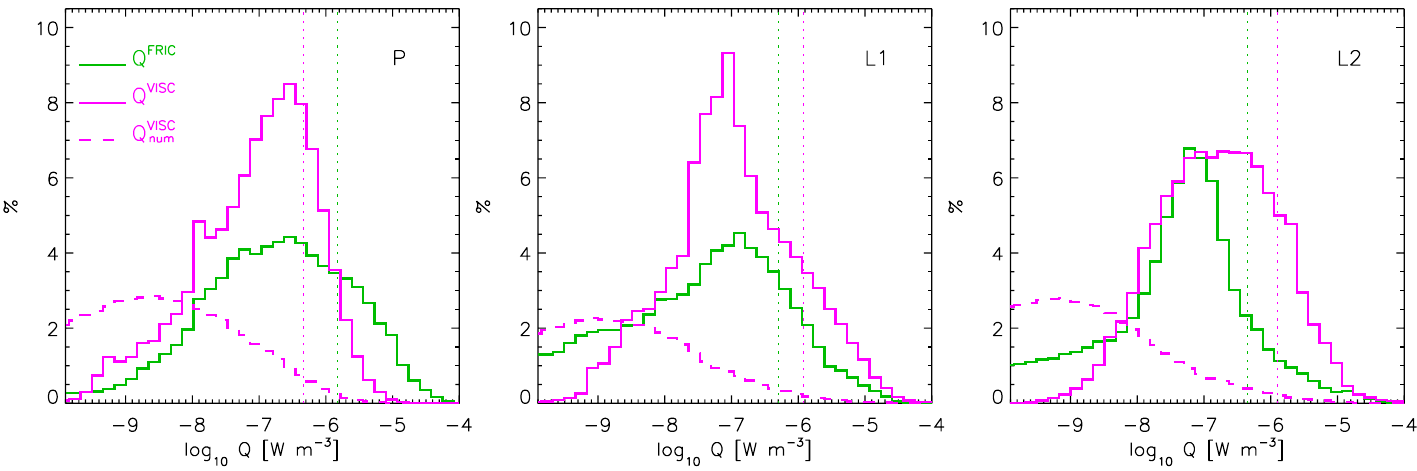}
\caption{Histograms of the frictional heating (green) and viscous heating (magenta) for the three magnetic configurations -- P (left), L1 (middle) and L2 (right) -- evaluated at the latest available simulations times corresponding to the right-most panels in Fig.~\ref{fig:RTIsnaps}. The dashed magenta line shows the artificial analogue of the viscous heating coming from the filtering. The vertical dotted lines mark the average values of the frictional and viscous heating.}
\label{fig:hist_qfric_qvisc}
\end{figure}

The dynamically evolving hot-cold plasma interfaces within the PCTR are sites characterized by both ion-neutral drifts and intense velocity gradients. These locations are expected to efficiently facilitate the operation of both frictional and viscous heating mechanisms. In Figure~\ref{fig:hist_qfric_qvisc}, histograms depict the frictional heating, $Q^{\rm FRIC}$ (green), the viscous heating, $Q^{\rm VISC}=Q^{\rm VISC}_{\rm n} + Q^{\rm VISC}_{\rm c}$ (solid purple), and the numerical viscous heating $Q^{\rm VISC}_{\rm num}$ (dashed purple) for each of the magnetic configurations evaluated at the same times as in Fig.~\ref{fig:corr}. Comparing physical and artificial viscous heating confirms that the latter is not a significant contributor to flow dissipation in our simulations, affirming the adequate spatial resolution of the flow structure. However, we note that the tail of the $Q^{\rm VISC}_{\rm num}$ distribution partly overlaps with the range of the $Q^{\rm VISC}$ distribution. We interpret this as evidence that the spatial resolution of approximately 1~km is the essential minimum required to resolve the underlying physics.

The histograms of $Q^{\rm FRIC}$ and $Q^{\rm VISC}$ in Fig.~\ref{fig:hist_qfric_qvisc} demonstrate the same change in the relative contribution of frictional and viscous heating from the P to L1 and L2 configurations using a statistical approach as shown with the integral measures in Fig.~\ref{fig:QvTime}. In the P simulation, $Q^{\rm FRIC}$ dominates, as is consistent with Fig.~\ref{fig:QvTime} and the considerably larger drift velocities observed in the P case in Fig.~\ref{fig:hist_dec}. One reason for this behaviour may be the more intricate flow structure observed when the magnetic field is strictly normal to the mixing plane. Consequently, the length of the hot-cold interface is notably greater in the P case compared to the L1 and L2 configurations.  On the other hand, there is a substantially stronger $Q^{\rm VISC}$ signal at the high $Q$ values in the L1 and L2 simulations.  One possible explanation of this behavior is that the interactions between plasma flows and in-plane magnetic fields may facilitate faster conversion of kinetic energy from large to small scales, leading to faster viscous dissipation.  An example of such conversion would be the magnetic reconnection process, where plasma inflow on large scales can be rapidly converted into plasma outflow through narrow funnel-like regions leading to large flow shear and consequent viscous dissipation.

\vspace*{-5pt}

\section{Conclusion summary}
This paper concludes a series of two-fluid studies of the linear and non-linear properties on the magnetized Rayleigh-Taylor instability in the context of a solar prominence \citep{rti1,rti2,rti3}.  We bring together, extend, and cross-compare several RTI simulations using three different magnetic field configurations within a two-dimensional numerical model.

We chose to include the discussion and interpretation of the presented results within the sections above, so here we only briefly summarize the key conclusions and observations:
\begin{itemize}
    \item 
    The RTI development in a plane normal to the direction of the magnetic field leads to fine-scale mixing between the cold prominence and hotter coronal material, which in turn leads to stronger ion-neutral flow drifts and stronger frictional heating than observed in sheared magnetic field configurations.  However, even in that case the magnitude of the kinetic energy contained in the relative drift between charges and neutrals is five orders of magnitude smaller than the total kinetic energy of the flows, i.e. the flows are well-coupled.

\item
    The nonlinear RTI energy release is fastest in the sheared magnetic field configuration with the slowest linear instability growth rate. The magnetic field within the mixing plane can lead to faster non-linear release of the gravitational energy driving the RTI, while more efficiently heating the plasma via viscous dissipation of associated plasma flows.
    While RTI evolution produces more than twice frictional than viscous heating for the P case, the viscous heating dominates over the frictional heating by approximately the same factor for the L1 and L2 configurations.  One possible explanation of this behavior is that the interactions between plasma flows and in-plane magnetic fields may facilitate faster conversion, through, e.g., magnetic reconnection,  of kinetic energy from large to small scales, leading to faster viscous dissipation. 
\item
    The time-integrated frictional heating is at all times almost exactly compensated by the time-integrated compressional cooling. The frictional heating and compressional cooling  act in equal and opposite directions with slight displacement, but generally in the same locations. This is consistent with the strong spatial correlation between the locations of large inverse neutral pressure gradient scale and the locations of strong frictional heating.

\item
    With the most substantial drifts concentrated in narrow layers near the hot-cold plasma interface, observationally integrating over under-resolved two-fluid sub-structure may lead to artificially inflated values of the drift velocities.

\end{itemize}

The results we have described here and in prior works may pose more questions for future investigations than provide clear-cut answers.  What is clear, however, is that more detailed fully three-dimensional two-fluid simulations of the magnetized RTI under realistic conditions are needed to provide accurate and robust interpetations of the current and future solar observations.  At the same time, better understanding of the fundamentals of RTI and similar mixing instabilities in a multi-fluid environment with multiple transport mechanisms can have benefits much beyond the solar and astrophysical communities.  To enable that, there is a need for even more idealized systematic studies exploring the effects of varying collisionality, magnetization, radiation transport, and kinetic plasma effects. 

\section*{Acknowledgement}

This work was supported by the Spanish Ministry of Science through the project PID2021-127487NB-I00 by the US National Science Foundation,  by the FWO grant 1232122N, and by the International Space Science Institute (ISSI) in Bern, through ISSI International Team project 457: The Role of Partial Ionization in the Formation, Dynamics and Stability of Solar Prominences It contributes to the deliverable identified in FP7 European Research Council grant agreement ERC-2017-CoG771310-PI2FA for the project “Partial Ionization: Two-fluid Approach”. The author(s) wish to acknowledge the contribution of Teide High-Performance Computing facilities to the results of this research. TeideHPC facilities are provided by the Instituto Tecnológico y de Energías Renovables (ITER, SA). URL: http://teidehpc.iter.es. Any opinion, findings, and conclusions or recommendations expressed in this material are those of the authors and do not necessarily reflect the views of the US National Science
Foundation.

\bibliography{aajour,biblio}

\begin{thebibliography}{19}
\providecommand{\natexlab}[1]{#1}
\providecommand{\url}[1]{\texttt{#1}}
\expandafter\ifx\csname urlstyle\endcsname\relax
  \providecommand{\doi}[1]{doi: #1}\else
  \providecommand{\doi}{doi: \begingroup \urlstyle{rm}\Url}\fi

\bibitem[{Anan} et~al.(2017){Anan}, {Ichimoto}, and {Hillier}]{2017Anan}
T.~{Anan}, K.~{Ichimoto}, and A.~{Hillier}.
\newblock {Differences between Doppler velocities of ions and neutral atoms in a solar prominence}.
\newblock \emph{A\&A}, 601:\penalty0 A103, May 2017.
\newblock \doi{10.1051/0004-6361/201629979}.

\bibitem[{Gonz{\'a}lez Manrique} et~al.(2023){Gonz{\'a}lez Manrique}, {Khomenko}, {Collados}, {Kuckein}, {Felipe}, and {G{\"o}m{\"o}ry}]{GonzalezManrique2023}
S.~J. {Gonz{\'a}lez Manrique}, E.~{Khomenko}, M.~{Collados}, C.~{Kuckein}, T.~{Felipe}, and P.~{G{\"o}m{\"o}ry}.
\newblock {Two fluid dynamics in solar prominences}.
\newblock \emph{arXiv e-prints}, art. arXiv:2311.03183, November 2023.
\newblock \doi{10.48550/arXiv.2311.03183}.

\bibitem[{Hillier} and {Arregui}(2019)]{Hillier+Arregui2019}
Andrew {Hillier} and I{\~n}igo {Arregui}.
\newblock {Coronal Cooling as a Result of Mixing by the Nonlinear Kelvin-Helmholtz Instability}.
\newblock \emph{ApJ}, 885\penalty0 (2):\penalty0 101, November 2019.
\newblock \doi{10.3847/1538-4357/ab4795}.

\bibitem[{Hillier} et~al.(2023){Hillier}, {Snow}, and {Arregui}]{Hillier+etal2023}
Andrew {Hillier}, Ben {Snow}, and I{\~n}igo {Arregui}.
\newblock {The role of cooling induced by mixing in the mass and energy cycles of the solar atmosphere}.
\newblock \emph{MNRAS}, 520\penalty0 (2):\penalty0 1738--1747, April 2023.
\newblock \doi{10.1093/mnras/stad234}.

\bibitem[{Khomenko} et~al.(2016){Khomenko}, {Collados}, and {D{\'{\i}}az}]{2016Khomenko}
E.~{Khomenko}, M.~{Collados}, and A.~J. {D{\'{\i}}az}.
\newblock {Observational Detection of Drift Velocity between Ionized and Neutral Species in Solar Prominences}.
\newblock \emph{ApJ}, 823:\penalty0 132, June 2016.
\newblock \doi{10.3847/0004-637X/823/2/132}.

\bibitem[{Leake} et~al.(2014){Leake}, {DeVore}, {Thayer}, {Burns}, {Crowley}, {Gilbert}, {Huba}, {Krall}, {Linton}, {Lukin}, and {Wang}]{Leake2014}
J.~E. {Leake}, C.~R. {DeVore}, J.~P. {Thayer}, A.~G. {Burns}, G.~{Crowley}, H.~R. {Gilbert}, J.~D. {Huba}, J.~{Krall}, M.~G. {Linton}, V.~S. {Lukin}, and W.~{Wang}.
\newblock {Ionized Plasma and Neutral Gas Coupling in the Sun's Chromosphere and Earth's Ionosphere/Thermosphere}.
\newblock \emph{Space Sci.\ Rev.}, 184:\penalty0 107--172, November 2014.
\newblock \doi{10.1007/s11214-014-0103-1}.

\bibitem[{Mart{\'\i}nez-G{\'o}mez} et~al.(2022){Mart{\'\i}nez-G{\'o}mez}, {Oliver}, {Khomenko}, and {Collados}]{MartinezGomez+etal2022}
David {Mart{\'\i}nez-G{\'o}mez}, Ram{\'o}n {Oliver}, Elena {Khomenko}, and Manuel {Collados}.
\newblock {Large Ion-neutral Drift Velocities and Plasma Heating in Partially Ionized Coronal Rain Blobs}.
\newblock \emph{ApJ}, 940\penalty0 (2):\penalty0 L47, December 2022.
\newblock \doi{10.3847/2041-8213/aca0a1}.

\bibitem[Meier and Shumlak(2012)]{2012Meier}
E.~T. Meier and U.~Shumlak.
\newblock A general nonlinear fluid model for reacting plasma-neutral mixtures.
\newblock \emph{Physics of Plasmas}, 19\penalty0 (7):\penalty0 072508, 2012.
\newblock \doi{10.1063/1.4736975}.
\newblock URL \url{http://dx.doi.org/10.1063/1.4736975}.

\bibitem[{Parenti} and {Vial}(2007)]{Parenti+Vial2007}
S.~{Parenti} and J.~C. {Vial}.
\newblock {Prominence and quiet-Sun plasma parameters derived from FUV spectral emission}.
\newblock \emph{A\&A}, 469\penalty0 (3):\penalty0 1109--1115, July 2007.
\newblock \doi{10.1051/0004-6361:20077196}.

\bibitem[{Parenti}(2014)]{Parenti2014}
Susanna {Parenti}.
\newblock {Solar Prominences: Observations}.
\newblock \emph{Living Reviews in Solar Physics}, 11\penalty0 (1):\penalty0 1, March 2014.
\newblock \doi{10.12942/lrsp-2014-1}.

\bibitem[{Popescu Braileanu} et~al.(2019){Popescu Braileanu}, {Lukin}, {Khomenko}, and {de Vicente}]{Popescu+etal2018}
B.~{Popescu Braileanu}, V.~S. {Lukin}, E.~{Khomenko}, and {\'A}.~{de Vicente}.
\newblock {Two-fluid simulations of waves in the solar chromosphere. I. Numerical code verification}.
\newblock \emph{A\&A}, 627:\penalty0 A25, Jul 2019.
\newblock \doi{10.1051/0004-6361/201834154}.

\bibitem[{Popescu Braileanu} et~al.(2021{\natexlab{a}}){Popescu Braileanu}, {Lukin}, {Khomenko}, and {de Vicente}]{rti1}
B.~{Popescu Braileanu}, V.~S. {Lukin}, E.~{Khomenko}, and {\'A}.~{de Vicente}.
\newblock {Two-fluid simulations of Rayleigh-Taylor instability in a magnetized solar prominence thread. I. Effects of prominence magnetization and mass loading}.
\newblock \emph{A\&A}, 646:\penalty0 A93, February 2021{\natexlab{a}}.
\newblock \doi{10.1051/0004-6361/202039053}.

\bibitem[{Popescu Braileanu} et~al.(2021{\natexlab{b}}){Popescu Braileanu}, {Lukin}, {Khomenko}, and {de Vicente}]{rti2}
B.~{Popescu Braileanu}, V.~S. {Lukin}, E.~{Khomenko}, and {\'A}.~{de Vicente}.
\newblock {Two-fluid simulations of Rayleigh-Taylor instability in a magnetized solar prominence thread. II. Effects of collisionality}.
\newblock \emph{A\&A}, 650:\penalty0 A181, June 2021{\natexlab{b}}.
\newblock \doi{10.1051/0004-6361/202140425}.

\bibitem[{Popescu Braileanu} et~al.(2023){Popescu Braileanu}, {Lukin}, and {Khomenko}]{rti3}
B.~{Popescu Braileanu}, V.~S. {Lukin}, and E.~{Khomenko}.
\newblock {Magnetic field amplification and structure formation by the Rayleigh-Taylor instability}.
\newblock \emph{A\&A}, 670:\penalty0 A31, February 2023.
\newblock \doi{10.1051/0004-6361/202142996}.

\bibitem[{Stellmacher} et~al.(2003){Stellmacher}, {Wiehr}, and {Dammasch}]{Stellmacher+etal2003}
G.~{Stellmacher}, E.~{Wiehr}, and I.~E. {Dammasch}.
\newblock {Spectroscopy of Solar Prominences Simultaneously From Space and Ground}.
\newblock \emph{Solar Phys.}, 217\penalty0 (1):\penalty0 133--155, October 2003.
\newblock \doi{10.1023/A:1027310303994}.

\bibitem[{Stone} and {Gardiner}(2007)]{Stone2007b}
James~M. {Stone} and Thomas {Gardiner}.
\newblock {The Magnetic Rayleigh-Taylor Instability in Three Dimensions}.
\newblock \emph{ApJ}, 671\penalty0 (2):\penalty0 1726--1735, Dec 2007.
\newblock \doi{10.1086/523099}.

\bibitem[Wiehr et~al.(2019)Wiehr, Stellmacher, and Bianda]{Wiehr2019}
E.~Wiehr, G.~Stellmacher, and M.~Bianda.
\newblock Evidence for the two-fluid scenario in solar prominences.
\newblock \emph{ApJ}, 873\penalty0 (2):\penalty0 125, mar 2019.
\newblock \doi{10.3847/1538-4357/ab04a4}.

\bibitem[{Wiehr} et~al.(2021){Wiehr}, {Stellmacher}, {Balthasar}, and {Bianda}]{Wiehr2021}
E.~{Wiehr}, G.~{Stellmacher}, H.~{Balthasar}, and M.~{Bianda}.
\newblock {Velocity Difference of Ions and Neutrals in Solar Prominences}.
\newblock \emph{ApJ}, 920\penalty0 (1):\penalty0 47, October 2021.
\newblock \doi{10.3847/1538-4357/ac1791}.

\bibitem[{Zapi{\'o}r} et~al.(2022){Zapi{\'o}r}, {Heinzel}, and {Khomenko}]{Zapior2022}
Maciej {Zapi{\'o}r}, Petr {Heinzel}, and Elena {Khomenko}.
\newblock {Doppler-velocity Drifts Detected in a Solar Prominence}.
\newblock \emph{ApJ}, 934\penalty0 (1):\penalty0 16, July 2022.
\newblock \doi{10.3847/1538-4357/ac778a}.

\end{thebibliography}

\end{document}